\documentclass{aa} 

\usepackage[dvipsnames]{xcolor}
\usepackage{amssymb}
\usepackage{natbib}
\usepackage{graphicx}
\usepackage{txfonts}
\usepackage[normalem]{ulem}
\usepackage{float}
\usepackage[caption = false]{subfig}

\begin{document}

\sloppy
  \title{Evolution of long-lived globular cluster stars}
   \subtitle{III. Effect of the initial helium spread on the position of stars in a synthetic Hertzsprung-Russell diagram	\thanks{The files containing the relevant evolution characteristics of the complete grid of models from the pre-main sequence up to the end of the stellar life \citep[see Appendix of][]{Chantereau15} are available in electronic form at the CDS via anonymous ftp to cdsarc.u-strasbg.fr (130.79.128.5) or via http://cdsweb.u-strasbg.fr/cgi-bin/qcat?J/A+A/. As in \cite{Chantereau15}, we also provide all the tables  on the website http://obswww.unige.ch/Recherche/evol/starevol/Globular.php.}} 

\authorrunning{W. Chantereau et al.} \titlerunning{III. Predictions of helium content in GCs}   

   \author{W. Chantereau \inst{1}
     \fnmsep\thanks{E-mail: william.chantereau@unige.ch}
          \and C. Charbonnel \inst{1,2} 
          \and G. Meynet \inst{1}
        }

   \institute{Department of Astronomy, University of Geneva, Chemin des Maillettes 51, CH-1290 Versoix, Switzerland
         \and
         IRAP, UMR 5277 CNRS and Universit\'e de Toulouse, 14 Av. E. Belin, F-31400 Toulouse, France}
   
  \date{}
 
  \abstract
  {Globular clusters host multiple populations of long-lived low-mass stars whose origin remains an open question.
   Several scenarios have been proposed to explain the associated photometric and spectroscopic peculiarities. They differ, for
instance, in the maximum helium enrichment they predict for stars of the second population, which these stars can inherit at birth as the result of the internal pollution of the cluster by different types of stars of the first population.}
  {We present the distribution of helium-rich stars in present-day globular clusters as it is expected in the original framework of the fast-rotating massive stars scenario (FRMS) as first-population polluters. We focus on NGC~6752.}
  {We completed a grid of 330 stellar evolution models for globular cluster low-mass stars computed with different initial chemical compositions corresponding to the predictions of the original FRMS scenario for [Fe/H]~=~-1.75. Starting from the initial helium-sodium relation that allows reproducing the currently observed distribution of sodium in NGC~6752, we deduce the helium distribution expected in that cluster at ages equal to 9 and 13 Gyr. We distinguish the stars that are moderately enriched in helium from those that are very helium-rich (initial helium mass fraction below and above 0.4, respectively), and compare the predictions of the FRMS framework with other scenarios for globular cluster enrichment.}
  {The effect of helium enrichment on the stellar lifetime and evolution reduces the total number of very helium-rich stars that remain in the cluster at 9 and 13~Gyr to only 12~\% and 10~\%, respectively, from an initial fraction of 21\%. Within this age range, most of the stars still burn their hydrogen in their core, which widens the MS band significantly in effective temperature. The fraction of very helium-rich stars drops in the more advanced evolution phases, where the associated spread in effective temperature strongly decreases. These stars even disappear from the horizontal branch and the asymptotic giant branch at 13~Gyr.}
  {The helium constraint is no suitable criterion for clearly
distinguishing between the scenarios for GC self-enrichment because only few very helium-rich stars are predicted in the investigated framework and because it is difficult to derive the helium content of GC stars observationally. However, the helium constraint indicates some difficulties of the original FRMS scenario that require the exploration of alternatives.}


        \maketitle
\section{Introduction}\label{intro}

Globular clusters (hereafter GCs) are among the oldest structures in the Universe. These compact stellar systems are classically referred to as the ideal laboratories for studying the evolution of low-mass stars, in particular because they have long been thought to host single age and chemically homogeneous stellar populations. However, several pieces of evidence have accumulated over the past two decades that indicate multiple (at least two) stellar populations in every individual GC studied in detail so far \citep[see, e.g., the review by][]{Gratton12}. 

High-resolution photometry reveals a wide 
variety of features 
in different areas of the color-magnitude diagram (CMD) of many GCs that are not consistent with single stellar populations \citep[see, e.g., the review of][]{Piotto09}.
Depending on the cluster, wide or multiple main sequences are observed \citep[MS,][]{Milone12d,Milone12e,King12}, multiple subgiant branches \citep[SGB,][]{Anderson09,Milone09b,Moretti09}, distinct red giant branches \citep[RGB,][]{Han09,Roh11,Monelli13}, or extended horizontal branches \citep[HB, e.g., ][]{Momany04,Busso07,Dalessandro11}.

Moreover, and although most individual GCs are rather homogeneous in terms of  Fe-peak element abundances \citep[e.g., ][]{Suntzeff93,James04,Carretta09a}, numerous spectroscopic studies 
have revealed significant  star-to-star variations in the abundances of light elements from carbon to magnesium. These ubiquitous variations are present in the form of anticorrelations (C-N, O-Na, Mg-Al; e.g., \citealt{Suntzeff91,Gratton01,Lind09,Carretta10}) that are the direct signatures of hydrogen-burning at high temperature \citep[e.g.,][]{DenisenkovD90,Prantzos07}. The O-Na anticorrelation is 
now considered as the typical characteristic of a \textup{\textup{\textup{{\it \textup{bona fide}}} }}GC \citep{Carretta10}. This pattern has only been observed in GCs so far and not in other environments such as open clusters \citep[see, for instance,][]{Bragaglia14,MacLean15}. We
note, however, that a very low fraction (1.5 to 4 \%) of halo stars that bear similar chemical features might have escaped from GCs \citep{Carretta10,Martell10,Schaerer11,Ramirez12}.
Importantly, the proportion of GC stars that present Na-enrichment and O-depletion with respect to field stars is very similar from one cluster to another ($\sim$ 50 to 80 \%; \citealt{Prantzos06,Carretta09b,Carretta10}).

These photometric and spectroscopic peculiarities are most commonly interpreted to result from the early pollution of the pristine intra-cluster matter by a first population (1P, often referred to as first generation) of short-lived stars that have long since
disappeared. They are called polluters and lead to the formation of a second population (2P) of anomalous stars  (with respect to the initial chemical composition of the proto-cluster and to that of field halo stars) that we observe today.
However, the exact processes governing the formation of multiple populations in GCs are far from being understood and are strongly debated (e.g., \citealt{BastianLardo2015,Renzinietal15,Krause16}). Competing scenarios invoke different types of possible 1P polluters, namely asymptotic giant branch stars (AGB; \citealt{Ventura01,D'Ercole10,Ventura11,Ventura13}), fast-rotating massive stars (FRMS;  \citealt{Prantzos06,Decressin07a,Decressin07b,Krause13,Chantereau15}), massive binary stars (\citealt{DeMink09}), supermassive stars (\citealt{Denissenkov14}), or a combination of some of the above-mentioned polluters (\citealt{Sills10,Bastian13}).

The most commonly invoked scenarios (AGB and FRMS) make very different predictions, in particular concerning the extent of helium enrichment that is associated with the O and Na abundance anomalies in 2P stars. 
While the FRMS scenario predicts that 2P stars can start their life with an helium mass fraction of between that of 1P stars (i.e., typically 0.248 for the metallicity presented in this study) and 0.8, the maximum helium enrichment provided by the AGB scenario amounts to at most $\sim$0.36-0.38 in mass fraction \citep[e.g.,][]{Siess10,Doherty14}. 
Since the initial helium content is an important ingredient for the evolution of stars (e.g., \citealt{IbenRood1969,Demarque71,Sweigart78,Maeder09,Chantereau15}), 
it is mandatory to quantify the effect of these differences on the expected properties of GC multiple populations (e.g., \citealt{D'Antona02,Salaris06_paper,Pietrinferni09,Sbordone11,Valcarce12,Cassisi13}).

In this series of papers, we explore the multiple consequences of the helium enrichment predicted by the FRMS in its original form for various properties of GCs and their host populations. 
\citet[][hereafter Paper I]{Chantereau15} discussed the implications of the assumed initial chemical distribution, and in particular of the initial helium abundance, on the characteristics, lifetimes, evolutional behavior, and fate of 2P low-mass stars. We also provided the ranges in initial mass and initial helium content of the stars that populate the different regions of the Hertzsprung-Russell diagram (HRD) at ages within the range covered by Galactic GCs. 
\citet[][hereafter Paper II]{Charbonnel16} thoroughly discussed the range in Na abundances along the AGB that is predicted for GCs of different metallicities and ages within the FRMS framework (see also \citealt{Charbonnel13}).

Here we study the distribution and number ratios of helium-rich stars in the different regions of the HRD of present-day GCs as predicted in the framework of the FRMS scenario, by means of population synthesis models. We extend the grid of stellar models presented in Paper I  (Sect.~\ref{framework}) with a high 
helium content and for a metallicity close to that of one of the best-studied GCs (in terms of chemical properties), namely NGC~6752 \citep[{[}Fe/H{]} = -1.56,]{Carretta09a,Villanova09}. We use these stellar models to build isochrones (Sect.~\ref{isochrones}) as well as population synthesis models (Sect.~\ref{section:syntheticGC}) of GCs based on simple assumptions on the IMF and on the distribution of the helium and sodium content of 1P and 2P stars at birth. 
These tools provide the distribution and number ratios of helium-rich and very helium-rich stars along the evolution sequence in the HRD at different ages (Sect.~\ref{section:resultsCMDs}). We discuss the results with respect to available observational constraints and describe the current state of the FRMS scenario (Sect.~\ref{Discussion}).
Finally, we summarize our conclusions and discuss the implications with respect to observations and to scenarios for GC enrichment (Sect.~\ref{Conclusions}).

\section{Stellar models}\label{framework}

\subsection{The grid}
We presented in Paper I a grid of 224 
stellar models for 1P and 2P GC low-mass stars (from 0.3 up to 1.0 M$_{\odot}$) that we computed with the stellar evolution code STAREVOL from the pre-main sequence up to the tip of the AGB phase (with some cases up to the planetary nebula phase). 
For the purpose of the present paper we extended the computations of Paper I to build a finer grid in mass and He abundance to assemble a more accurate tool for the population synthesis calculations. 
The complete grid is now composed of 330 stellar models with the  metallicity $Z = 5.4 \times 10^{-4}$, initial masses M$_\mathrm{ini}$/M$_{\odot}$ = [0.3, 0.35, 0.4, 0.45, 0.47, 0.5, 0.55, 0.6, 0.65, 0.67, 0.7, 0.75, 0.8, 0.9, 1.0], and initial helium mass fraction Y$_\mathrm{ini}$ = [0.248, 0.26, 0.27, 0.3, 0.33, 0.37, 0.4, 0.425, 0.45, 0.475, 0.5, 0.525, 0.55, 0.575, 0.6, 0.625, 0.65, 0.675, 0.7, 0.73, 0.77, 0.8]. 
These standard models do not include atomic diffusion, rotation-induced mixing, or overshooting. However, mass-loss is accounted for throughout the evolution following the prescriptions of \citet[][with $\eta$ = 0.5]{Reimers75} and \cite{Vassiliadis93} before and after the end of central helium burning, respectively. 

\begin{figure}[h]
   \centering
   \includegraphics[width=0.47\textwidth]{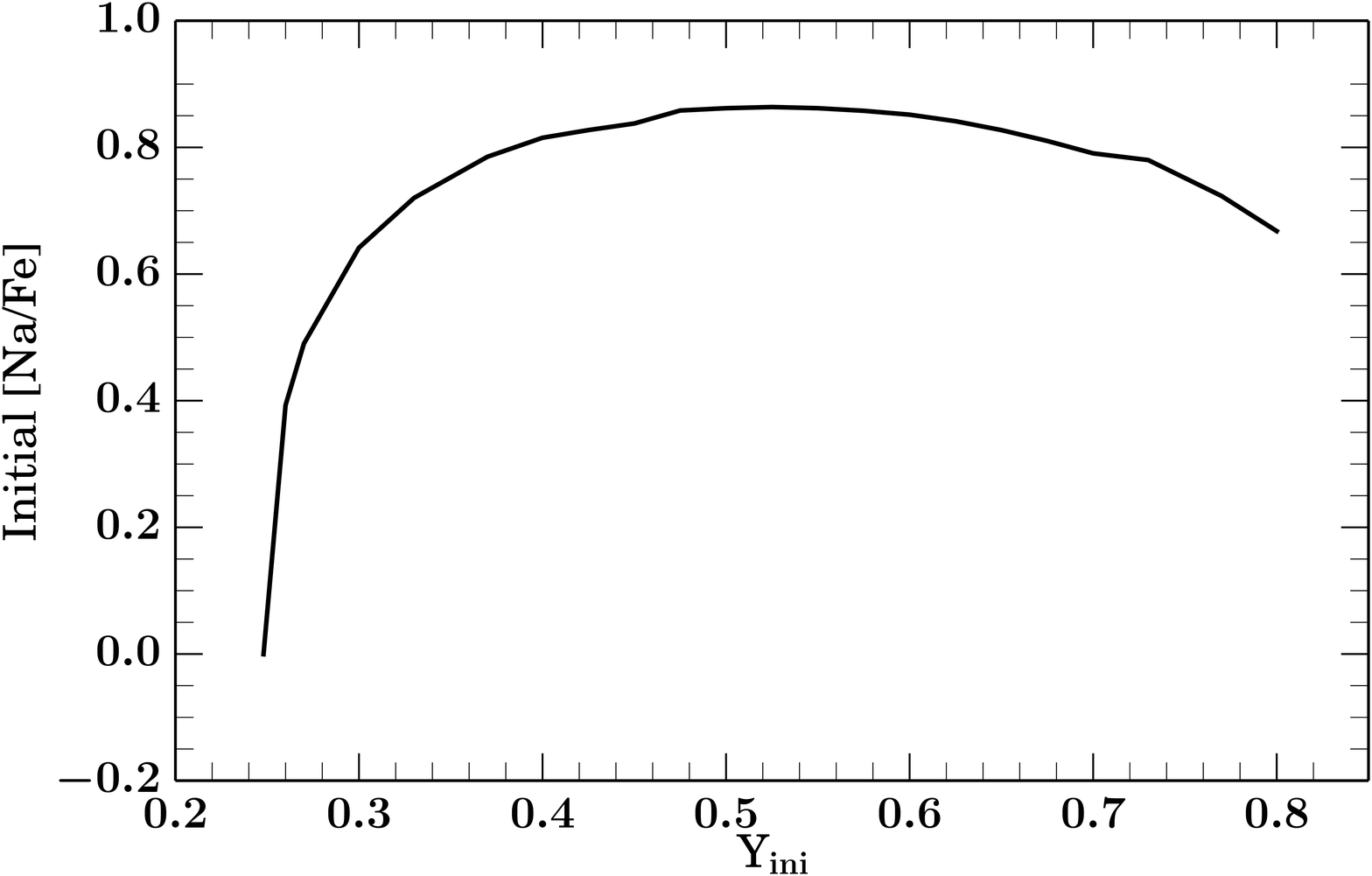}
    \caption{Theoretical distribution for the initial sodium abundances ([Na/Fe] = log($N_\mathrm{Na}/N_\mathrm{Fe})_*$-log($N_\mathrm{Na}/N_\mathrm{Fe})_\odot$) as a function of the initial helium mass fraction adopted for our 2P stars models following \citet{Decressin07a} for the dilution between pristine gas and FRMS ejecta.}
    \label{NaFe_FRMS}
\end{figure} 

\begin{figure}[h]
   \centering
   \includegraphics[width=.47\textwidth]{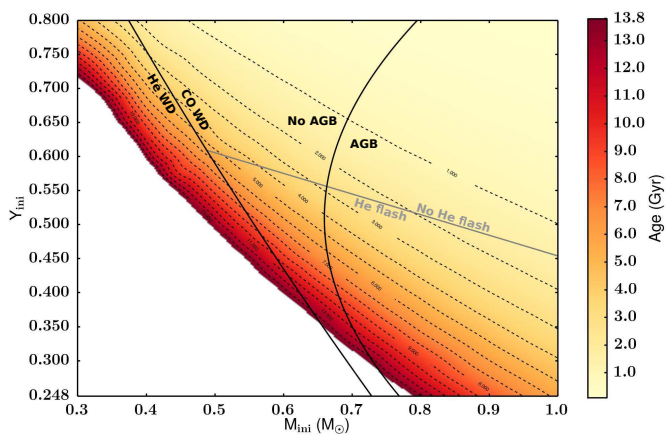}
    \caption{Diagram representing the age of the stars at the turnoff (color-coded) as a function of the initial helium mass fraction and mass for the grid of 330 models ($Z = 5.4 \times 10^{-4}$). The dashed lines link the stellar models reaching the turnoff at a given age (numbers, in Gyr),  and the white area corresponds to turnoff ages later than 13.8~Gyr. The thick black lines separate the different behaviors of the models; the left line delimits the domains where stars become either He or CO white dwarfs, and the right line separates the stars that climb the AGB from those that do not. The gray line delimits the domain where helium ignition (when it occurs) starts with a flash or in non-degenerate conditions.}
  \label{figure:diagram}
\end{figure}

\subsection{Initial chemical composition}
\label{icc}

All the models were computed for [Fe/H]~=~-1.75 ($Z = 5.4 \times 10^{-4}$). We  considered an $\alpha$-enrichment of +0.3~dex typical for Galactic GCs \citep[e.g.,][]{Carretta10}. This corresponds to the values used by \citet{Decressin07a} for their FRMS models from which we derived the He-Na relation for our 2P stellar models (see below, and Paper~I). The spectroscopic determination of [Fe/H] for NGC~6752 (-1.56$\pm$0.03; \citealt{Carretta09a,Villanova09}) is slightly higher than the value we use, but this does not qualitatively affect our conclusions, which mainly depend on the assumed He-Na relation for 2P stars. However, the numbers we give for the current masses, effective temperature, luminosities, and numbers of stars in specific regions of the HRD at various ages (Sect. \ref{isochrones} and \ref{section:resultsCMDs}),
for instance, would be slightly different quantitatively if we had computed the models for the  observed value of [Fe/H]. 

We assumed that the chemical composition of the first stellar population reflects that of the original proto-cluster material.
For this metallicity, we adopted an initial helium mass fraction Y$_\mathrm{ini}$ of 0.248 and an initial [Na/Fe] equal to 0 for the 1P GC stars. This [Na/Fe] value agrees with the mean value of 0.01$\pm$0.03 derived by \citet{Carretta12} for the so-called primordial population in NGC~6752, and it is slightly higher than the lowest [Na/Fe] value of -0.14$\pm$0.05 observed in NGC~6752 by \citet{Carretta13}. This small shift has no effect on the conclusions of this paper because the exact content of Na does not affect stellar evolution \citep[e.g.,][]{Pietrinferni09,VandenBerg12}. However, we compensate for the corresponding offset when we compare the theoretical and observed Na distributions  (Sect.~\ref{InitialHedistribution}).

With respect to their 1P counterparts, 
our stellar models for 2P stars are initially depleted in C, O, Mg, Li, Be, and B, and enriched in He, N, Na, and Al to various degrees.
The overall C+N+O content is held constant in the two populations
because no He-burning products are included in 2P stars, as predicted by the FRMS models and in agreement with observations \citep{Smith96,Carretta05}. 
We assumed a range in the initial mixture for 2P stars that results from the dilution of the H-burning products ejected by 1P FRMS models with intra-cluster pristine gas following the assumptions of \citet{Decressin07a} to reproduce the Li-Na anticorrelation observed in NGC~6752 (see details in Sect.~3.2.2 of Paper I).

We show in Fig.~\ref{NaFe_FRMS} the corresponding initial relation between the abundances of He and Na of our grid of stellar models (see also Fig.~1 of Paper I).
The highest initial mass fraction in He, Y$_\mathrm{ini}$, is 0.8. In the following, we distinguish between stars that are moderately enriched in helium (Y$_\mathrm{ini}$ below or equal to 0.4) and those that are very helium rich (Y$_\mathrm{ini}$ above 0.4). The initial theoretical range in [Na/Fe] covers 0.86~dex. This is slightly lower than the spread of 0.98~dex derived by \citet{Carretta13} in NGC~6752. This difference is a direct consequence of the assumed dilution factor between pristine gas and FRMS raw ejecta, in which [Na/Fe] reaches typical values between 1.0 and 1.5~dex for a 60~M${\odot}$ star. Changing the hypothesis on the dilution factor would therefore easily help to reconcile the extent of the assumed and observed initial spread in Na. However, we decided not to play with this quantity as this would not affect our general conclusions (see also Sect.~3.2.3 of Paper I). 

\section{Effect of initial helium on the stellar models and on the isochrones}\label{isochrones}

We quantified in Paper I how the variations in initial helium mass fraction over the considered range affect the duration of the main evolutionary phases and modify the evolutionary paths in the HRD for the whole grid of models. The main results are summarized in Fig.~\ref{figure:diagram}, where we also indicate the mode of He-ignition when it occurs (We take the presence or absence of the characteristic bend in the log($\rho_\mathrm{c}$)/log(T$_\mathrm{c}$) plane at the end of the RGB phase as the limit between He-flash/no He-flash behaviors, see, e.g., \citealt{Chantereau15}, Fig.~5), the nature of the white dwarf remnants (He- or CO-WD), and the ability for a star to climb  the AGB depending on its initial mass and helium content.

\begin{figure}
   \centering
   \includegraphics[width=.47\textwidth]{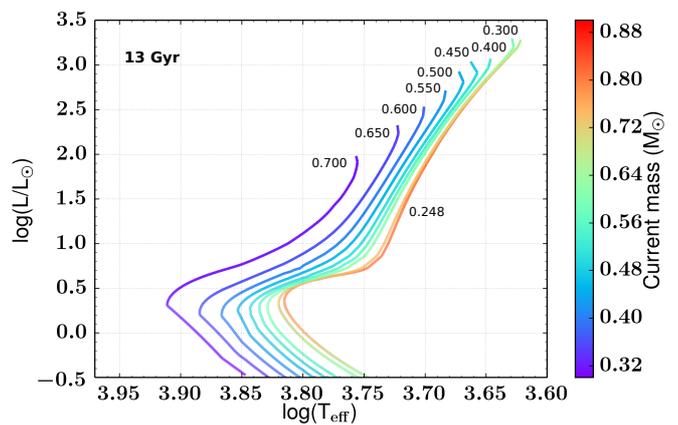}
    \caption{Theoretical isochrones at 13~Gyr for different initial helium mass fractions Y$_\mathrm{ini}$ from 0.248 up to 0.7 (labels) and abundances of other elements varying accordingly (see Paper I for details). The color code represents the current stellar mass (in M$_{\odot}$) at 13~Gyr.} 
  \label{figure:isochrones}
\end{figure}

\begin{table*}[ht]
        \centering 
        \begin{tabular}{c | c|c || c|c|c || c|c || c|c|c}       
        \hline 
        \multicolumn{11}{c}{} \\
        \multicolumn{11}{c}{\textbf{9~Gyr}} \\
        \multicolumn{11}{c}{} \\
        \hline
         Y$_\mathrm{ini}$ &  \multicolumn{2}{c}{L$_{\odot}$} & \multicolumn{3}{c}{L$_\mathrm{turnoff}$} & \multicolumn{2}{c}{L$_{1.5}$} & \multicolumn{3}{c}{L$_\mathrm{RGB tip}$} \\ \hline
        & log(T$_{\rm eff}$) & M (M$_{\odot}$) & log(T$_{\rm eff}$) & M (M$_{\odot}$) & log$\left(\frac{L}{L_{\odot}}\right)$ & log(T$_{\rm eff}$) & M (M$_{\odot}$) & log(T$_{\rm eff}$) & M (M$_{\odot}$) & log$\left(\frac{L}{L_{\odot}}\right)$ \\ \hline    
    0.248 &   3.80 & 0.78 & 3.84 & 0.87 & 0.50 & 3.72 & 0.90 & 3.62 & 0.73 & 3.32 \\
        0.300 &   3.81 & 0.73 & 3.84 & 0.80 & 0.50 & 3.72 & 0.82 & 3.62 & 0.65 & 3.30 \\
        0.400 &   3.82 & 0.62 & 3.85 & 0.67 & 0.43 & 3.72 & 0.68 & 3.64 & 0.54 & 3.24 \\
        0.500 &   3.84 & 0.52 & 3.86 & 0.55 & 0.41 & 3.73 & 0.56 & 3.66 & 0.45 & 3.11 \\
        0.600 &   3.86 & 0.43 & 3.89 & 0.44 & 0.36 & 3.74 & 0.45 & 3.69 & 0.40 & 2.80 \\
        0.700 &   3.89 & 0.34 & 3.93 & 0.35 & 0.39 & 3.77 & 0.35 & 3.74 & 0.34 & 2.35 \\
    
        \hline 
        \multicolumn{11}{c}{} \\
        \multicolumn{11}{c}{\textbf{13~Gyr}} \\
        \multicolumn{11}{c}{} \\
        \hline
        0.248 & 3.80 & 0.74 & 3.82 & 0.79 & 0.36 & 3.71 & 0.81 & 3.62 & 0.63 & 3.28 \\
        0.300 & 3.80 & 0.69 & 3.82 & 0.72 & 0.33 & 3.72 & 0.74 & 3.63 & 0.55 & 3.30 \\
        0.400 & 3.82 & 0.58 & 3.83 & 0.60 & 0.28 & 3.72 & 0.61 & 3.65 & 0.50 & 3.07 \\
        0.500 & 3.83 & 0.49 & 3.84 & 0.50 & 0.27 & 3.73 & 0.50 & 3.67 & 0.43 & 2.93 \\
        0.600 & 3.86 & 0.39 & 3.87 & 0.40 & 0.20 & 3.74 & 0.40 & 3.70 & 0.37 & 2.53 \\
        0.700 & 3.89 & 0.31 & 3.91 & 0.32 & 0.31 & 3.77 & 0.32 & 3.76 & 0.31 & 1.98 \\       
        \hline
        \end{tabular} 
\caption{T$_{\rm eff}$, M (M$_{\odot}$), luminosity at log$\left(\frac{L}{L_{\odot}}\right)$ = 0, at the turnoff, at log$\left(\frac{L}{L_{\odot}}\right)$ = 1.5 (along the RGB), and at the RGB tip for 9 and 13~Gyr.} 
\label{table:isochrones}
\end{table*}

Compared to He-normal stars and for a given initial mass, helium-enriched stars evolve faster, they are brighter and hotter on the main sequence and the subgiant branch, and they ignite central helium burning at lower luminosity on the red giant branch (see Paper I for details).
This directly affects the positions of the isochrones and the current stellar mass along a given isochrone when
 different initial helium mass fractions are assumed (see also \citealt{Valcarce12} and \citealt{Cassisi13_MmSAI}, but for lower He spread than considered here).

Examples are shown in Fig.~\ref{figure:isochrones}, where we focus on the evolution from the zero-age main sequence to the RGB tip in the HRD at 13 Gyr (i.e., representative of an old GC like NGC 6752, with 13.4$\pm$1.1~Gyr
according to \citealt{Gratton03}). 
In Table~\ref{table:isochrones} we quantify the effects on the luminosity, effective temperature, and current stellar mass along the 9 and 13~Gyr isochrones with different helium enrichments at key evolution points (on the main sequence at the luminosity of the Sun, at the turnoff, on the RGB at log$\left(\frac{L}{L_{\odot}}\right)$ = 1.5, and at the RGB tip).
This variety of behaviors is expected to affect the properties of the stellar populations in different areas of GC HRDs, as discussed below.  

\section{Building synthetic globular clusters}
\label{section:syntheticGC}

\subsection{General method}

We used a modified version of the population synthesis code Syclist \citep{Georgy14}
to follow the evolution of stellar populations as a function of time, 
taking stellar model predictions for different helium contents into account. 
We assumed that the first and second stellar populations were all born at the same time. This might not be the case in real globular clusters where different populations might form at different epochs. In the frame of the FRMS scenario, however, the expected time delay between the formation of the first and second populations is  at most 4.5 to 9 Myr \citep{Krause13}, which is less than 0.1~\% of typical GC age ($\sim$ 12.4 Gyr, \citealt{Carretta10}). In the AGB scenario this time delay is also predicted to be relatively short compared to the age of GCs, on the order of 90-120~Myr \citep[e.g.,][]{DAntona16}. Therefore, in both cases these very small shifts in age are entirely negligible for our conclusions.

We assumed the same initial stellar mass function for 1P and 2P stars. For stellar masses of up to 0.85 M$_{\odot}$, we used the present-day mass function derived by \cite{Paresce00}, which corresponds to a log-normal shape $\mathrm{ln}~\Phi(log(M)) = A - \frac{[log(M/M_c)]^2}{2\sigma^2}$ with $A$ a normalization constant, $\langle M_{c}\rangle = 0.33 \pm 0.03$ and $\langle\sigma\rangle = 0.34 \pm 0.04$. 
For more massive stars, we used a power-law distribution $\Phi(M) = \frac{dN}{dM} = B M^{-(1+x)}$ , where B is a normalization constant and $x = 1.35$ corresponding to the \cite{Salpeter55} value. We followed the evolution of an initial population of 300'000 stars with initial masses between 0.3 and 1 M$_\odot$. This large number is sufficient to avoid strong stochastic effects.  
We did not account for dynamical effects that might lead to the ejection of stars after their formation.
We also neglected the effects of close binary evolution (tidal interactions, mass transfer) since the number of binaries observed in GCs is quite low (lower than $\sim$3-5~\% in NGC~6752; \citealt{Milone12_binaries,Ji15}). 
Therefore, the number of nuclear active stars within the cluster evolves with time because of stellar evolution effects alone. We present our results for 9 and 13~Gyr old synthetic clusters, which covers the age spread derived for Galactic GCs \citep[e.g.,][]{MarinFranch09,VandenBerg13}.

\begin{figure*}[ht]
   \centering
\subfloat[]{\includegraphics[width = 3in]{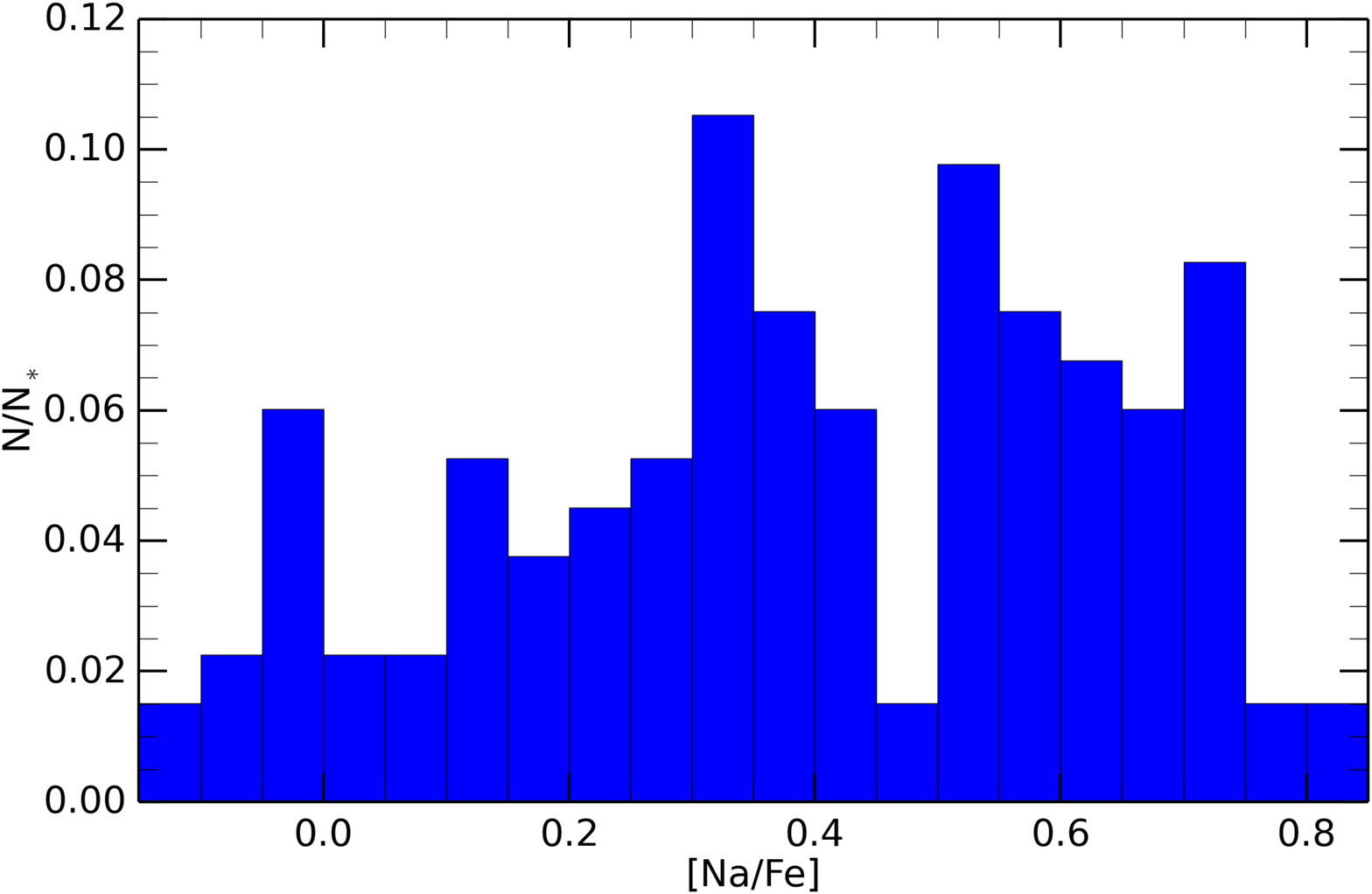}} 
\subfloat[]{\includegraphics[width = 3in]{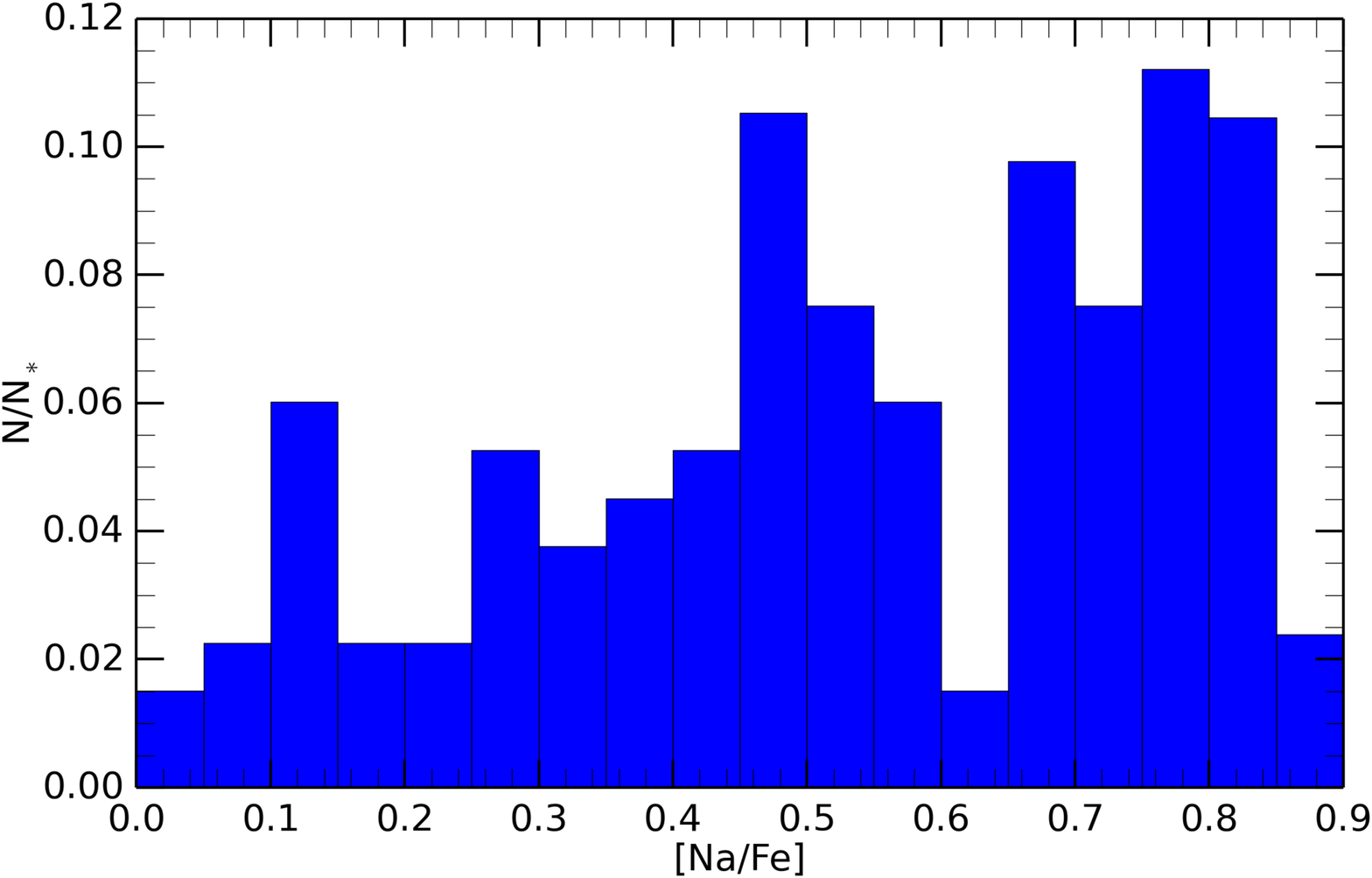}}\\
\subfloat[]{\includegraphics[width = 3in]{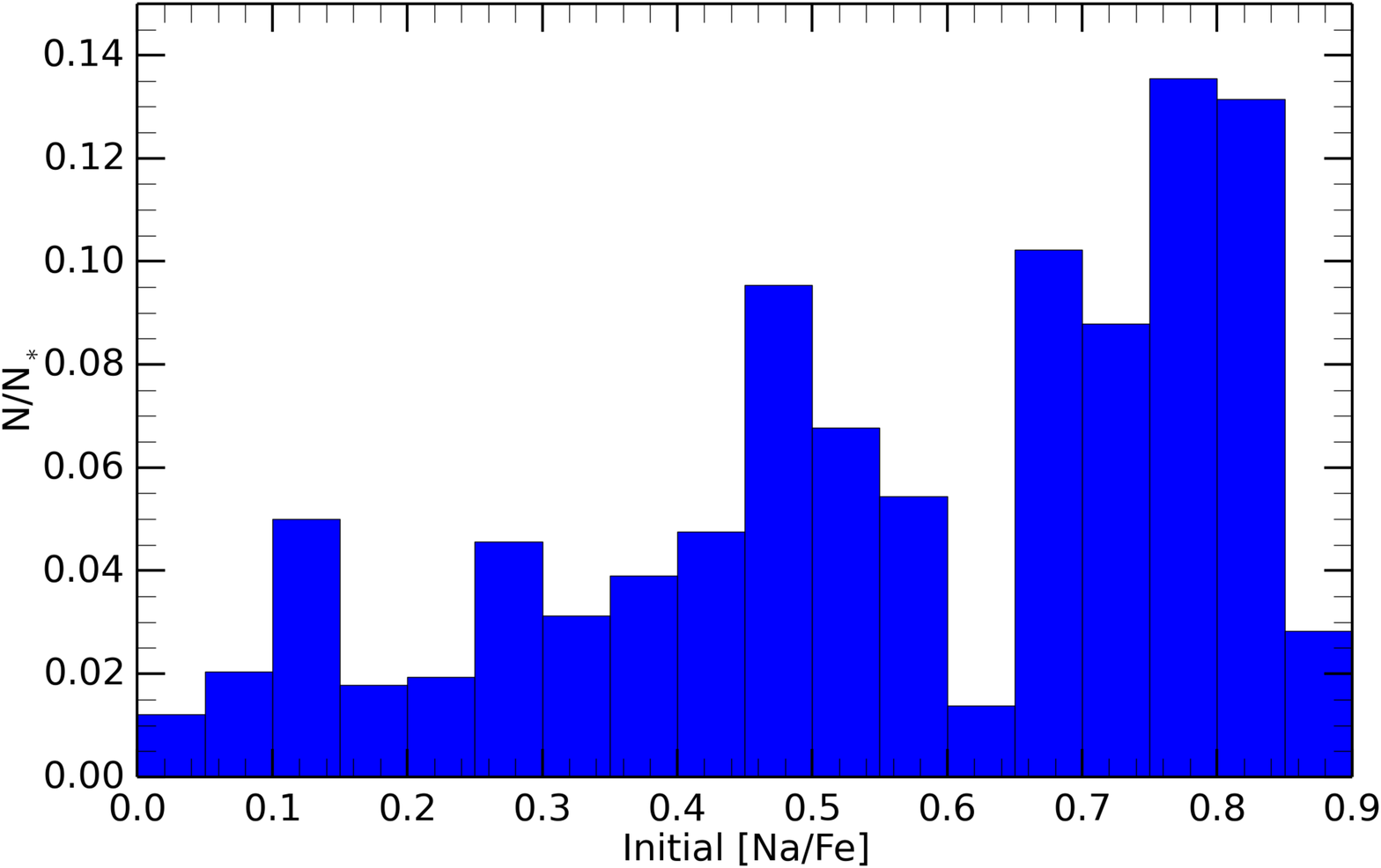}}
\subfloat[]{\includegraphics[width = 3in]{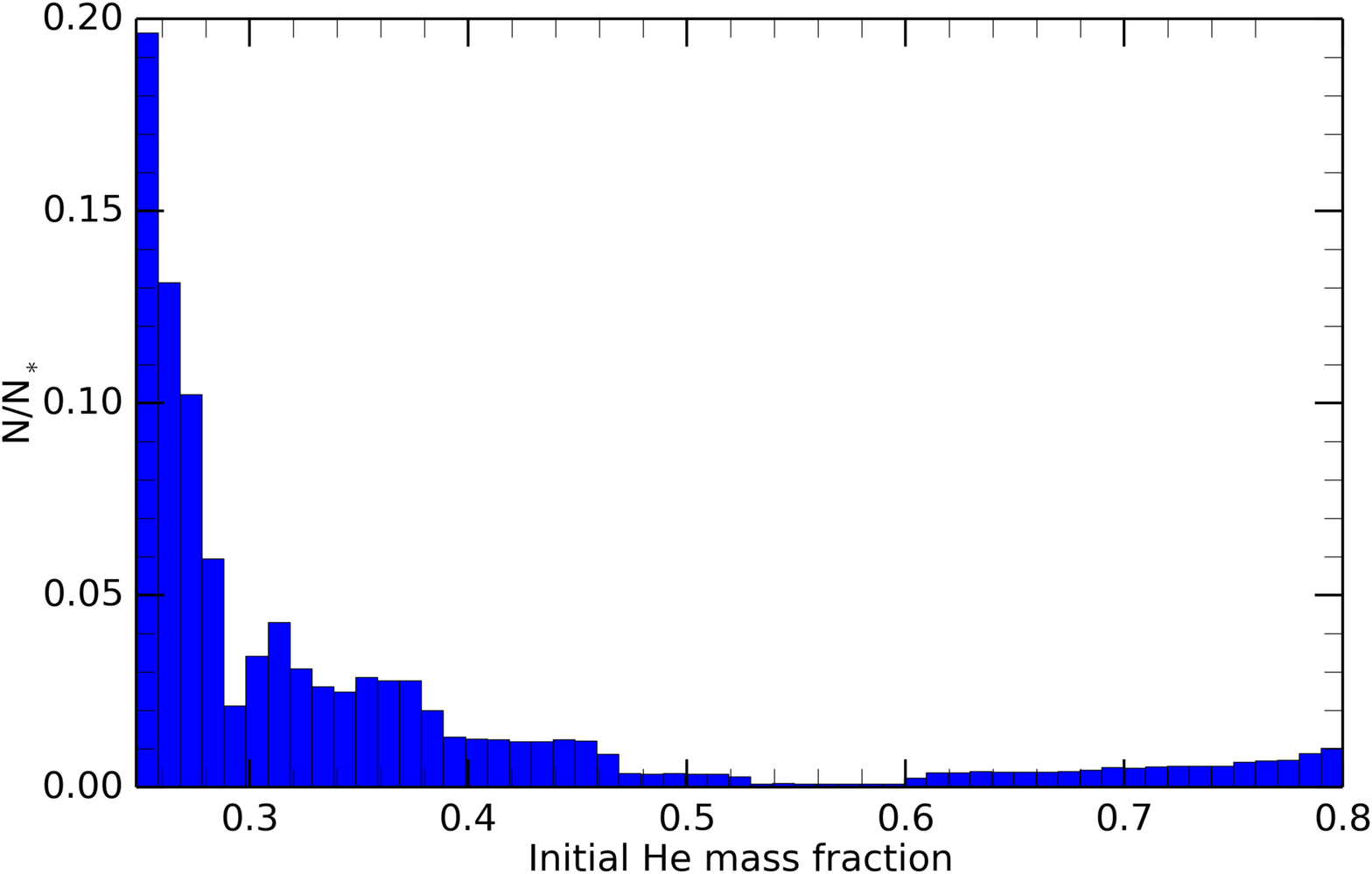}} 
    \caption{{\it (Panel a)} [Na/Fe] distribution for the 133 bright red giants observed  in NGC~6752 by \cite{Carretta13}. {\it (Panel b)}    
Data from \cite{Carretta13} presented in {\it panel a} adapted to be used in the FRMS framework. {\it (Panel c)} Initial [Na/Fe] distribution. {\it (Panel d)} Initial helium mass fraction inferred from the theoretical distribution for the initial sodium abundance in the framework of the FRMS scenario.}
    \label{Heini}
\end{figure*} 

\subsection{Initial distribution of stars along the theoretical He-Na relation}
\label{InitialHedistribution}

A key ingredient for building our synthetic stellar population is the initial distribution of the sample stars along the theoretical He-Na relation (i.e., how many stars have each given \{He;Na\} abundance). This 
must lead to a theoretical abundance distribution at 13 Gyr that reproduces the observed distribution of [Na/Fe] in NGC 6752, taking into account the evolution of the stars with different initial helium abundance and mass. According to standard stellar models such as those presented here, but also according
to more sophisticated models that include rotation and thermohaline mixing \citep[e.g.,][]{CZ07,Lagarde12}, the abundances of sodium (and obviously of iron) at the surface of red giants with masses of around 1~M$_\odot$ or lower are not modified by any process in the red giant stars themselves. This is consistent with the fact that the Na-O anticorrelation is observed at the MS turnoff and at different luminosities along the RGB. This excludes the
possibility that any additional very deep mixing acts in GC RGB stars, as initially invoked by \citet{WeissDenissenkovCC2000} before spectroscopic analyses of MS GC stars. 
We used the spectroscopic observations of 133 red giants in this GC by \citet[][and private communication]{Carretta13}; the corresponding [Na/Fe] distribution is shown in Fig.~\ref{Heini} (panel a). We shifted the whole observed distribution by 0.14~dex, which is the difference between the lowest [Na/Fe] value assumed in our stellar models and the lowest value observed by \citet[][see Sect.~\ref{icc}]{Carretta13}.
Additionally, and as explained in Sect.~\ref{icc}, 
the theoretical range of [Na/Fe] 
covered by our models extends over 0.86 dex, whereas the observations display a range of 0.98~dex. To take the stars into account that
display a $\Delta$[Na/Fe] between 0.86 and 0.98 in our study, we therefore split them in the bins between $\Delta$[Na/Fe] = 0.75 and 0.86. Given the uncertainties on [Na/Fe] abundances (typically $\pm$0.05~dex) and on the predictions of the models, it is consistent to shift these stars to lower [Na/Fe] bins. Moreover, these stars represent only a minor part of the GC population, $\sim$ 8 \% of the sample of \cite{Carretta13}. These adjustments lead to the [Na/Fe] distribution shown in Fig.~\ref{Heini} (panel b), which is the observational constraint that our synthetic population must reproduce at 13~Gyr.

We determined the initial [Na/Fe] distribution by an iterative process. We first assumed that it is equal to the distribution observed in the 13~Gyr cluster NGC~6752 (Fig.~\ref{Heini}, panel b), and we attributed an initial He mass fraction to the sample stars following the theoretical He-Na relation. Because stars of similar initial masses but  different initial compositions have different lifetimes, the theoretical [Na/Fe] distribution at 13~Gyr was modified with respect to the initial guess. To minimize the differences between the theoretical distribution at 13~Gyr and the distribution observed in NGC~6752, we then modified the initial distribution and performed a new iteration. We stopped the iteration process when the theoretical [Na/Fe] distribution at 13~Gyr reproduced the observations.

The initial distributions of [Na/Fe] and helium mass fraction that we derived and used for our synthetic GC models 
are shown in Fig. \ref{Heini} (panels c and d, respectively). The initial sodium distribution displays a greater number of high [Na/Fe] values than the observed distribution on the RGB at 13~Gyr (Fig.~\ref{Heini}, panel b), as expected.
This is directly linked to the fact that the most He- and Na-enriched stars die sooner than their less He-enriched counterparts, which decreases the fraction of stars with the highest [Na/Fe] values that are still alive at 13~Gyr. 
Importantly, the very He-enriched stars represent only a small fraction of the whole initial stellar population. While 49 \% of the stars display an initial He-enrichment above 0.3 in mass fraction, only 21 \% are expected to be born with an helium content higher than 0.4. We discuss the remaining percentages of these stars in synthetic populations at 9 and 13~Gyr below.

\begin{figure}[h]
   \centering
   \includegraphics[width=0.47\textwidth]{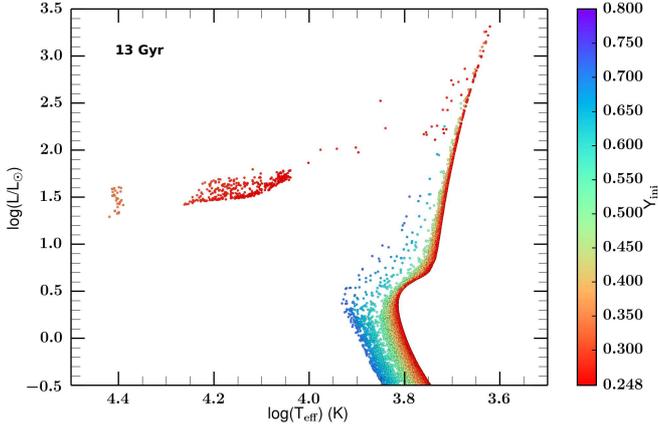}
    \caption{Position of the synthetic GC stars in the HRD at 13~Gyr. The color code corresponds to the initial helium mass fraction of the stars that are still alive at these ages (199'887 out of the 300'000 initial sample).}
    \label{HRD}
\end{figure} 

\section{Distribution of helium-rich and very helium-rich stars in a synthetic HRD}
\label{section:resultsCMDs}

\subsection{General results}
We built synthetic models of GCs at the ages of 9 and 13~Gyr by 
following the evolution of an initial sample of 300'000 stars based on our stellar models and assuming the initial helium and sodium abundance distributions and the stellar IMF  described above.
As a result of stellar evolution effects, only 225'501 (75 \% of the initial sample) and 199'887 stars (67 \%) are still alive (i.e., undergoing nuclear burning) at 9 and 13~Gyr, respectively.
The positions of the stars in the HRD at 13~Gyr are shown in Fig.~\ref{HRD}, where the color code corresponds to the initial helium mass fraction of individual stars. 

From this we directly obtain the theoretical  distribution of the stars at key locations within the HRD of present-day GCs as a function of their initial He content. Quantitative predictions for the number of stars born with different initial helium abundances that are present at 9 and 13~Gyr at different evolution phases are given in Table~\ref{table:distributions}.
We give the number ratios of stars with Y$_\mathrm{ini}$ above 0.3, 0.35, and 0.4 to pinpoint the expected differences between the AGB and the FRMS scenarios (the first allowing for initial helium up to $\sim$ 0.36 - 0.38, the second up to 0.8). We also give the highest value for the initial helium abundance Y$_\mathrm{ini, max}$ of the stars that lie at the different evolution phases within the FRMS framework.

\begin{table}[ht]
        \centering 
        \begin{tabular}{c c | c  c  c  | c }       
        \hline 
        Gyr & Phase & Y$_\mathrm{ini}$ & Y$_\mathrm{ini}$ & Y$_\mathrm{ini}$ & Y$_\mathrm{ini, max}$ \\ 
     & Phase & $\geq~0.3$ & $\geq~0.35$ & $\geq~0.4$ & \\ \hline
        \textbf{0} & - & 49 & 33 & 21 & 0.800 \\ 
        \hline
        & All & 41 & 24 & 12 & 0.763 \\
        & MS & 42 & 25 & 12 & 0.752 \\
        \textbf{9} & Lower RGB & 43 & 25 & 12 & 0.763 \\
        & Upper RGB & 42 & 26 & 12 & 0.763 \\
        & HB & 45 & 27 & 12 & 0.512 \\
        & AGB & 33 & 13 & 0 & 0.382 \\  
        \hline
        & All & 39 & 22 & 10 & 0.729 \\
        & MS & 39 & 22 & 11 & 0.724 \\
        \textbf{13} & Lower RGB & 46 & 25 & 13 & 0.729 \\
        & Upper RGB & 38 & 22 & 10 & 0.729 \\
        & HB & 37 & 12 & 0.4 & 0.407 \\
        & AGB & 10 & 0 & 0 & 0.305 \\
        \hline \\ 
        \end{tabular} 
\caption{Percentage of stars with initial helium mass fractions higher than 0.3, 0.35, and 0.4 that are present in the cluster at different evolution phases at 3 ages (0, i.e., initial values, 9, and 13~Gyr). 
``All" corresponds to all the cluster stars with  log$\left(\frac{L}{L_{\odot}}\right)$ above -0.5 (71'187 and 53'194 stars at 9 and 13~Gyr, respectively). ``MS" corresponds to main sequence stars with log$\left(\frac{L}{L_{\odot}}\right)$ between -0.5 and 0.3. ``Lower and upper RGB“ corresponds to red giant branch stars with luminosities between 0.8 and 1.2 on one hand and to  brighter stars up to the RGB tip on the other hand. ``HB" includes all the central He-burning stars and ``AGB" all the stars that are climbing the asymptotic giant branch. In the last column we give the highest value for the initial helium abundance of the stars in the different evolution phases.} 
\label{table:distributions}
\end{table}

\begin{figure}[ht!]
   \centering
   \includegraphics[width=0.5\textwidth]{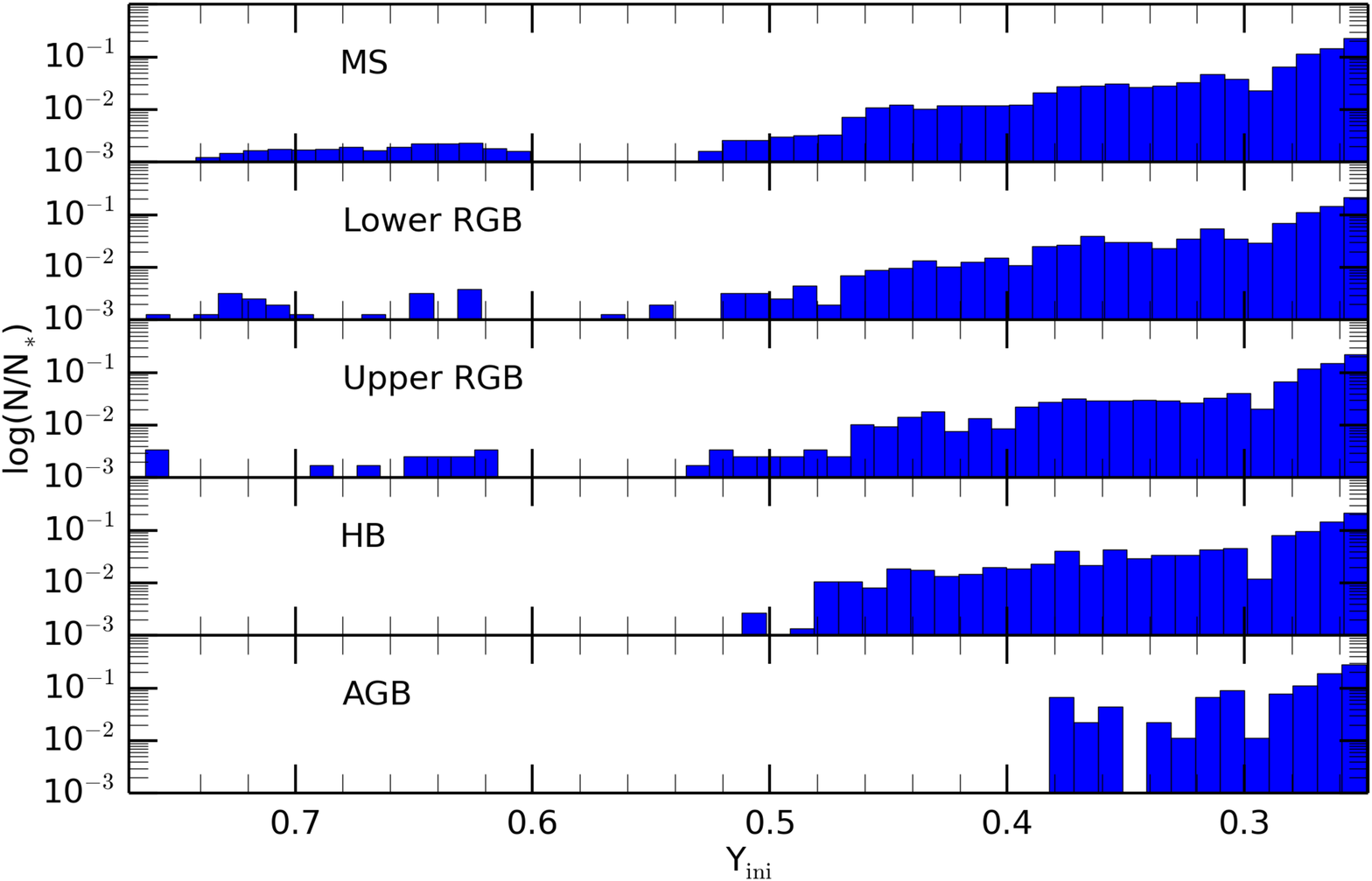}
   \includegraphics[width=0.5\textwidth]{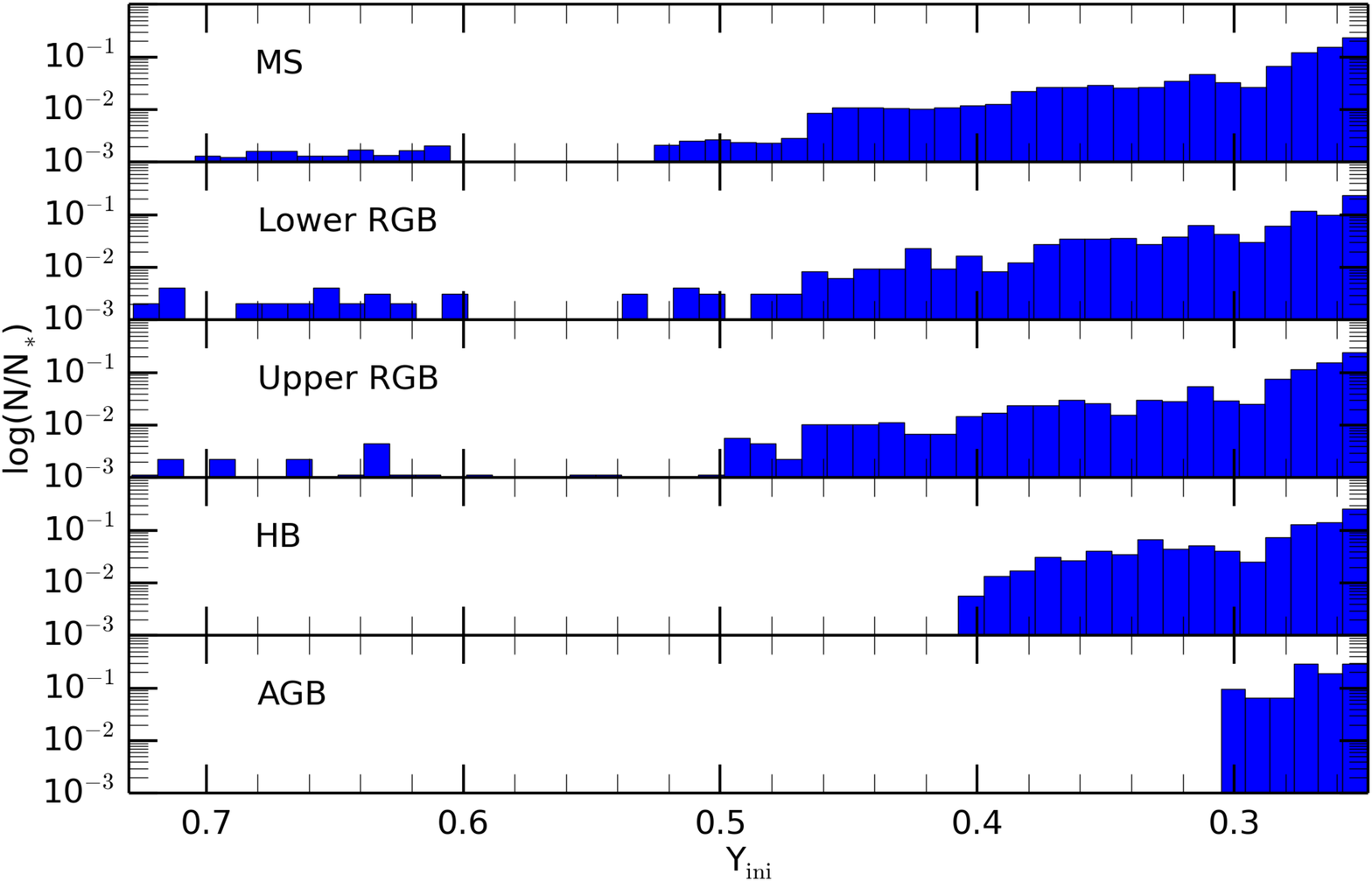}
    \caption{Distribution of stars as a function of their initial helium content (in mass fraction) at 9 and 13~Gyr (top and bottom, respectively) for the different evolution phases defined in Table~\ref{table:distributions}.}
\label{figure:distribution_9_13gyr}
\end{figure} 

\begin{figure}
    \centering
   \includegraphics[width=0.5\textwidth]{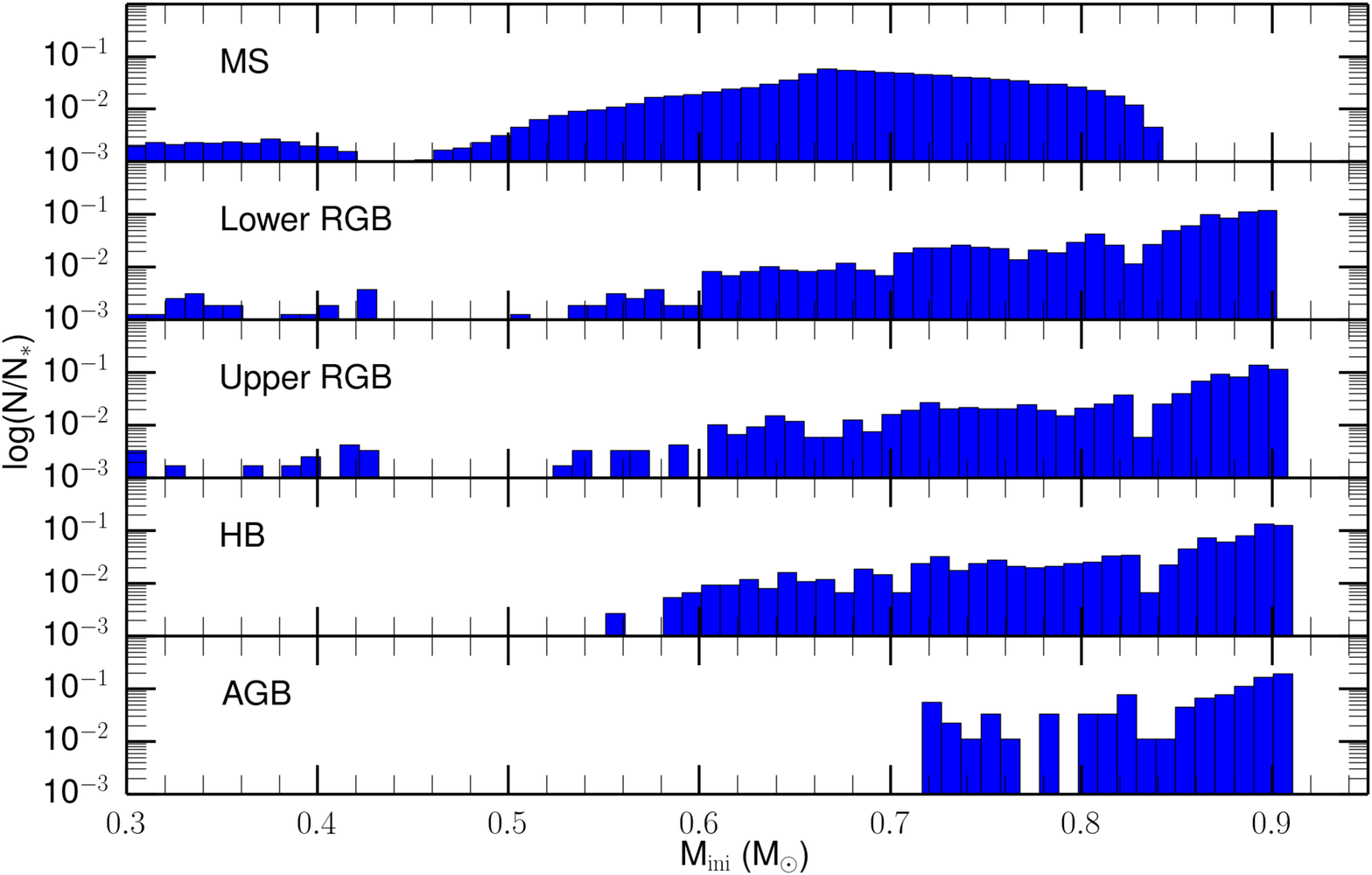}
   \includegraphics[width=0.5\textwidth]{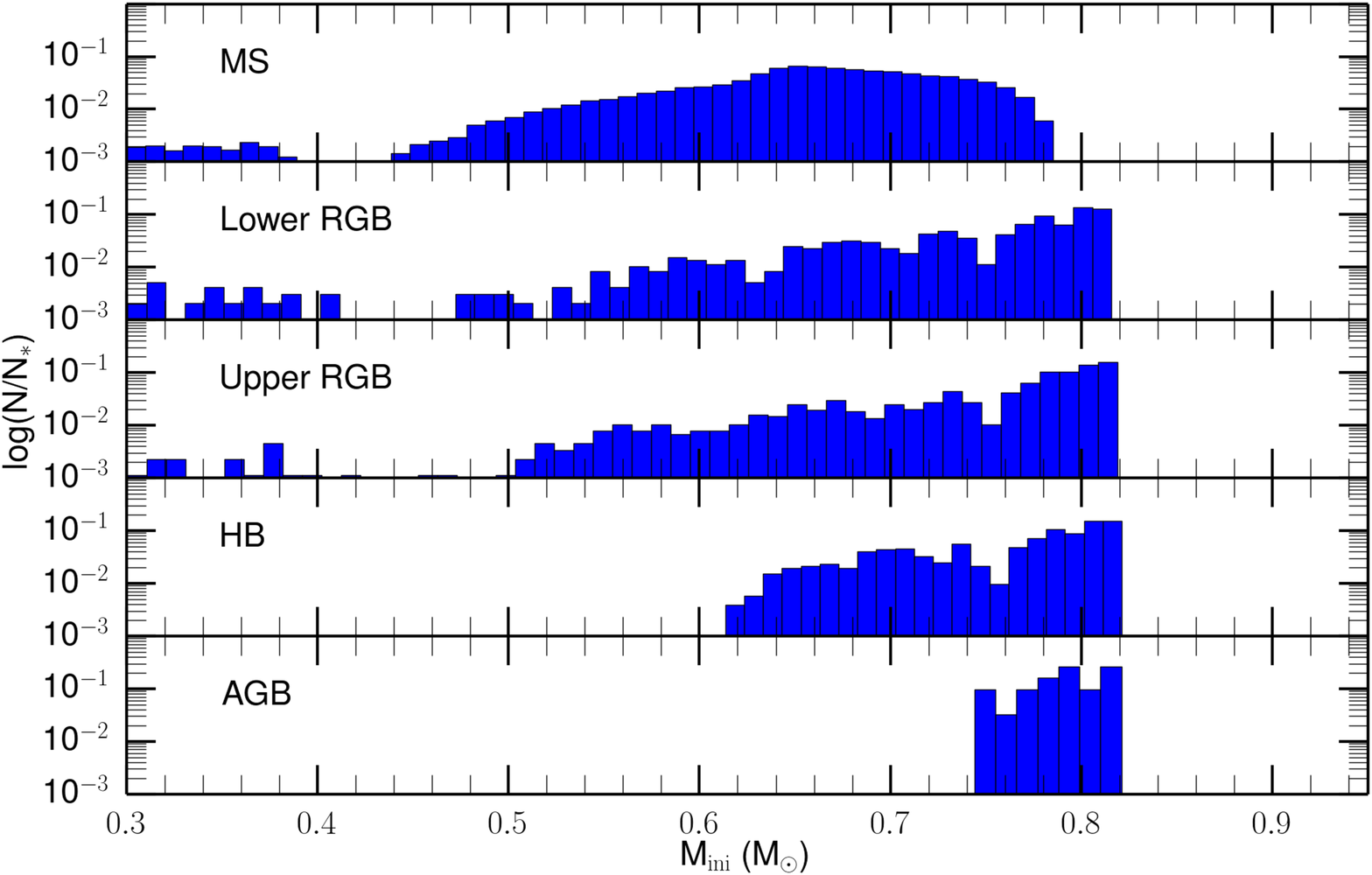}
     \caption{Distribution of stars as a function of their initial mass at 9 and 13~Gyr (top and bottom, respectively) for the different evolution phases defined in Table~\ref{table:distributions}.}
     \label{figure:distributionMini}
\end{figure}

The predicted distribution and number ratios of helium-rich stars change drastically with time through the effect of the initial helium content of stars on their evolution paths and lifetimes. While the percentage of stars with initial helium mass fraction Y$_\mathrm{ini}$ higher than 0.35 (0.4) is 33\% (21\%) within the original sample, only 24\% (12\%) remain at 9~Gyr and 22\% (10\%) at 13~Gyr. The contribution of very He-rich stars (specific to the FRMS scenario) to the overall synthetic population is thus relatively modest and slightly decreases with time. The results are summarized in Figs.~\ref{figure:distribution_9_13gyr} 
and \ref{figure:distributionMini}, where we show the distribution of stars with different initial helium content and corresponding initial masses for specific evolution phases that we discuss in more detail below.

\subsection{Main sequence}

The effect of helium enrichment on the stellar effective temperature scale, hence on stellar colors, causes the whole synthetic sample of stars to display a wide spread in effective temperature on the main sequence (Fig.~\ref{HRD}). At the turnoff (i.e., around log$\left(\frac{L}{L_{\odot}}\right) \sim 0.3$) this spread is on the order of 1820~K for 13~Gyr. The hottest turnoff star in our synthetic diagram has a Y$_\mathrm{ini, max}$ of 0.714 and an initial stellar mass of 0.311~M$_{\odot}$. These quantities are only slightly different (i.e., higher) for the 9~Gyr old synthetic cluster.

Importantly, at the MS turnoff MS at 13~Gyr, 29\% of stars have Y$_\mathrm{ini}$ between 0.3 and 0.4, and only 10 \% have Y$_\mathrm{ini}$ above 0.4.
Therefore, the effect on the observed broadening of the main sequence that is due to the relatively rare very He-rich stars that are specific to the FRMS framework is relatively modest (compare to the AGB scenario that does not predict stars with 
Y$_\mathrm{ini}$ higher than 0.36 - 0.38).

\subsection{Subgiant and red giant branches}

At 9 and 13~Gyr, the SGB and RGB are much narrower than the MS. During these evolution phases, stars with modest He enrichment (Y$_\mathrm{ini}\sim0.3$) have only slightly higher effective temperatures than 1P stars. 
The apparent broadening of the SGB and of the base of the RGB is due to 2P stars born with a higher helium content. 
Additionally, the distribution obtained for Y$_\mathrm{ini}$ along the RGB is very sensitive to the considered luminosity bin. For the reasons given when we discussed the effect of helium on the isochrones, the higher the luminosity on the RGB, the lower the largest predicted initial helium mass fraction Y$_\mathrm{ini, max}$ . 
At 13~Gyr, RGB stars with Y$_\mathrm{ini}$ higher than 0.4 account for only 13 and 10 \% of the stars on  the lower and upper RGB, respectively (i.e., for log$\left(\frac{L}{L_{\odot}}\right)$ below or above 1.2, respectively). Again, these  percentages are slightly higher at 9~Gyr.

\subsection{Horizontal branch}

\subsubsection{Limitations of the interpolation procedure}
Despite the refined discretization of our grid of stellar models, there are still artifacts in the advanced evolution phases caused
by the interpolation between models of stars that have sharply different behaviors. For instance, when we interpolate between models that burn helium on the horizontal branch (HB) and those that evolve directly toward the WD cooling curve after climbing the RGB, the synthesis code predicts stars in incorrect regions of the HRD. More precisely, the code places them above the actual luminosity of the HB over a wide range in effective temperatures.
Most of them should be located on the blue part of the HB, however,
and on the extreme HB (EHB) because of their helium content (0.3$<\mathrm{Y}_\mathrm{ini}<$0.41). 
These artifacts, incorrectly located stars, represent a non-negligible fraction of the total number of our sample of HB stars, namely 32~\%.  We decided to not show them in the HRDs of Fig.~\ref{HRD} to avoid confusion; as a consequence, the gap is expected to be less pronounced. 

\subsubsection{Highest helium content on the horizontal branch}
Despite the above-mentioned limitation and as already discussed in Papers I, II, and \citet{Charbonnel13}, we can conclude that only stars with  Y$_\mathrm{ini}$ lower than 0.41 populate the HB at 13~Gyr in our synthetic GC (see Fig.~\ref{figure:distribution_9_13gyr}). Moreover, only 9~\% (3~\%) of the HB stars have an initial helium mass fraction between 0.36 and 0.41 (between 0.38 and 0.41). 
It is important to recall that the AGB scenario predicts a maximum helium enrichment of 2P stars  of Y$_\mathrm{ini, max} \sim 0.36-0.38$ \citep{Siess10,Doherty14}. 
This means that the AGB and the FRMS scenario predict a similar maximum extension of the initial helium content on the HB for analogous ages and metallicities, hence it is impossible to use this evolution phase to distinguish between these two frameworks.

\subsection{Asymptotic giant branch}

As discussed extensively in Paper II and \cite{Charbonnel13}, the highest value for the initial helium content Y$_\mathrm{ini, max}$ of AGB stars is relatively low in the FRMS scenario because
of the AGB-manqu\'e behavior (see, e.g., the limit AGB/no-AGB for our grid of models in Fig.~\ref{figure:diagram}). Additionally, the number of He-rich stars dramatically drops with increasing cluster age (see Fig.~\ref{figure:distribution_9_13gyr}).
In the 31 AGB stars that are still present in our synthetic cluster at 13~Gyr (9~Gyr), Y$_\mathrm{ini, max}$ is 0.305 (0.382) and corresponds to a current stellar mass of 0.536~M$_{\odot}$, while the highest current mass found in this phase is 0.598~M$_{\odot}$ for the stars born with the lowest helium content. 

The absence of very He-rich stars that we predict along the AGB agrees with the observations of \cite{Campbell13}, who reported a lack of very Na-rich AGB stars in the old GC NGC~6752 (see also \citealt{Charbonnel13}, and \citealt{Wang16}). However, in contrast to Campbell and collaborators, who concluded that in this GC all 2P stars fail to climb the AGB, we obtain that $\sim$74~\% of the AGB stars at 13~Gyr belong to the 2P with the current model assumptions. \cite{Cassisi14} also found a non-negligible proportion of 2P stars on the AGB ($\sim$50~\%) thanks to synthetic horizontal branch simulations for NGC~6752. 
Theoretical works also emphasize the fact that the number of AGB stars in a given GC strongly depends on the assumptions made on the mass loss in the previous evolution phases \citep{Charbonnel13,Charbonnel16,Cassisi14}. Moreover, in the framework of the FRMS scenario, the theoretical Na spread on the AGB phase is a function of both age and metallicity \citep{Charbonnel16}, and the expected trends can explain the presence of 2P stars  found in most of the GCs studied so far,
which cover a range in age and iron content (47~Tuc, NGC~2808, M~3, M~5, M~13, M~55, M~2, see \citealt{Johnson12,GarciaH15,Johnson15}; and Wang et al. II in prep.) with the exception of M~62 \citep{Lapenna15}. 

\section{Discussion}\label{Discussion}

\subsection{Observational constraints}
\label{obs}

\subsubsection{Main sequence}
Helium differences between 1P and 2P stars have been indirectly derived from isochrone fitting of the spreads in the main-sequence colors in ultraviolet and optical filters for a handful of GCs.
A largest He difference of only $\sim +$0.03 (in mass fraction) between 1P and 2P stars  has been deduced with this method in NGC~6752 \citep{Milone13}. 
 If we consider that in this cluster Y$_\mathrm{ini}$ of 1P stars is 0.248 as assumed in the present study, this means that the largest helium mass fraction Y$_\mathrm{ini,max}$ of 2P stars is 0.278.

This extremely low helium enrichment is difficult to reconcile with the current FRMS scenario given that we predict that $\sim$ 50~\% of the stars have an initial helium content above this value at 13~Gyr. 

As of today, the largest helium spread on the MS determined for a GC is $\sim +$0.13 in NGC~2808, which hosts a triple MS \citep{Milone12d}. This corresponds to Y$_\mathrm{ini,max}$ of $\sim$0.378, assuming the same value for Y$_\mathrm{ini}$ than in NGC~6752, although this cluster is slightly more metal-poor and older than NGC~2808.
In our synthetic GC, $\sim 15 \%$ of the MS stars are predicted to be born with a higher He content. Whether or not this relatively modest proportion of stars could be identified through photometry and isochrone fitting should be investigated by observers in the future. We still consider this as a challenge for the original FRMS scenario, as discussed in more detail in Sect.~\ref{csquencesFRMS}. 

\subsubsection{Horizontal branch}
Direct measurements of He abundances in GC stars are extremely challenging. Non-local thermodynamic equilibrium analyses were performed  for a small sample of HB stars in NGC~2808 \citep{Marino14}. In this cluster the bluest HB stars that could be studied have effective temperature of $\sim$~11'000~K. They present evidence of an He enhancement of $+$0.09$\pm$0.06 with respect to the predicted primordial He content, which agrees relatively well with both the FRMS and the AGB self-enrichment scenarios.

However, the most He-enhanced stars are expected to populate the extreme blue tail of the HB at effective temperatures higher than $\sim$~18'000~K. Unfortunately, the effects of atomic diffusion (gravitational settling of He, radiative levitation of metals; \citealt{Hui-Bon-Hoa00}) on the surface He abundances are very strong there and reach a depletion factor of 300 in some cases (i.e., a derived He atmospheric abundance lower than the cosmological value; \citealt{Behretal00,Behr03}). Therefore, the He range predicted along the HB can hardly be used to constrain the different self-enrichment scenarios.

\subsection{Consequences for the FRMS scenario}\label{csquencesFRMS}

The FRMS scenario in its original form \citep{Decressin07a,Decressin07b} proposes that 2P stars form out of the material ejected by massive stars that rotate close to the critical velocity and that this
material  mixes with original proto-cluster gas. The input physics and the prescriptions that were assumed to build Decressin's FRMS models (in particular the treatment of mass loss at critical velocity) imply that the burning ashes of the CNO cycle and of the NeNa and MgAl chains are released by the polluters through slow equatorial winds throughout their evolution on the MS and the luminous blue variable phases. This explains the form of the He-Na relation shown in Fig.~\ref{NaFe_FRMS} and the high helium enrichment expected in a relatively large number of 2P stars considered in the present work. 

However, stars more massive than $\sim$ 20~M$_{ \odot}$  reach central temperatures high enough for the NeNa chain to operate fully already at the beginning of the main sequence \citep{Prantzos07}. In the 60~M$_{ \odot}$ FRMS model of \citet{Decressin07b}, the strongest Na increase is reached in the stellar core at a temperature of 49~MK when the central mass fraction of helium is only 0.262. We remind that these numbers are given for a 60~M$_{ \odot}$ FRMS model computed for [Fe/H] of -1.75. However, they are expected to vary with metallicity
because for a given stellar mass, changing the [Fe/H] ratio modifies the internal temperature of the star and affects the nucleosynthesis. 
The He-Na relation presented in this specific study would then
be different as well. This behavior is clearly depicted in Fig.~\ref{NaFe_FRMS}, which shows that very high Na-enrichment is reached for a helium mass fraction of about 0.3 (see also Fig.~1 of Paper I, where the behavior of all the elements involved in H-burning is shown for massive polluters of different masses).
Since MS massive stars essentially consist of a very extended convective core, $\sim 60 - 70 \%$ of the stellar mass very quickly reaches the correct chemical composition to explain the Na 
abundance patterns of 2P without too strong He enrichment (this does not include mixing by fast rotation, which changes
the original composition of the external radiative layers with H-burning products). 

The question is then how this matter can be ejected by massive stars early on in the main sequence before the Na-rich ejecta become contaminated by too much helium. Clearly, the mechanical mass loss at critical rotation and the radiatively driven winds are not sufficiently efficient with the prescriptions currently used to compute models of massive stars. Therefore, we have to call for additional or modified processes that would be unique to GC massive stellar hosts. We are currently investigating various instability mechanisms that may lead to the required mass-loss rates in specific stellar environments such as very dense proto-GCs. 

In the present paper we focused on the Na and He content of GC 2P stars. However, Mg-depletion is also observed in the most Na- and Al-rich stars in NGC~6752 \citep{Yong03}.  Higher central hydrogen burning temperatures of about 70-75~MK are required to significantly deplete  $^{24}$Mg through proton captures \citep[e.g.,][]{Prantzos07}. However, such high temperatures are reached in stars with masses between 60 and 120~M$_{\odot}$ (the highest polluter mass considered in the original FRMS scenario) only after the mid-MS, when He-enrichment is already non-negligible (see Fig.~1 of Paper~I). Mg depletion can be obtained earlier on in the MS in massive stars when we arbitrary increase the $^{24}$Mg(p,$\gamma$) reaction rate \citep{Decressin07b}. 
The same is true for stars with masses of 200 and 500~M$_{\odot}$ computed with $Z = 5 \times 10^{-4}$ with the Geneva evolution code (C.Georgy, private communication). 
Mg depletion in 2P stars is therefore expected to occur together with relatively high He-enrichment if FRMS are the polluters, unless we accept to adjust the nuclear reaction rates or consider stars more massive than $\sim$ 200 - 500~M$_{\odot}$ to be at the origin of the Mg-Na anticorrelation in GCs. In this context, hypothetical super-massive stars with masses higher than 10$^4$~M$_{\odot}$ might be the best candidate polluters, as initially suggested by \citet{Denissenkov14}. As clearly explained by \citet{Denissenkovetal15}, only these stars reach a central temperature of about 75~MK,
which is necessary to deplete Mg  with the current prescriptions for the nuclear reaction rates early on in the MS before strong He-enrichment occurs. Whether stars like this can exist in very unique environments such as infant GCs, and how their ashes might
be recycled into a second population of stars remains to be understood. 

We showed in this paper that the original FRMS scenario predicts a widening of the MS band.  This is certainly a weakness, since observations show distinct sequences in most GCs.  
A possible solution to this difficulty in the frame of this model might be linked to the conditions for low-mass star formation that may only be realized when peculiar abundances are reached. In the present form of the scenario, low-mass stars can form at any moment, while in reality this is probably not true.

Finally, all GCs present large differences in the abundance spreads of H-burning products (e.g., the extent of the O-Na anticorrelation varies from one cluster to another) and in the He spreads derived from isochrone fitting. 
This clearly poses a problem for all the polluters proposed in the literature, at least within the current frameworks and for the available yields.  We agree on this point with \citet{Bastianetal15}, who conclude that high degrees of stochasticity are necessary to explain the broad variety of chemical patterns observed in GCs.

\section{Summary}\label{Conclusions}

We studied the effect of the initial helium abundances predicted for 1P and 2P GC stars within the original FRMS scenario on the distribution and number ratios of helium-rich stars at different evolution phases in the theoretical HRD.
We focused on a relatively metal-poor GC ([Fe/H]=-1.75). We used the corresponding FRMS yields of \citet{Decressin07b} and the [Na/Fe] distribution observed by \citet{Carretta13} for NGC~6752  to reconstruct the distribution of the initial composition (and in particular the initial sodium-helium distribution) of
a sample of 300'000 GC low-mass stars for which we assumed a standard IMF. Based on our large grid of stellar evolution models constructed with the suitable chemical composition, we used our population synthesis tool to follow the evolution of this large stellar sample over time.
We did not consider stellar evaporation from the cluster over its lifetime, but as a result of stellar evolution, the number of stars still alive (i.e., undergoing nuclear burning)  decreased to 225'501 and 199'887 at 9 and 13~Gyr, respectively.   

A direct comparison of our theoretical predictions with observed CMDs is beyond the scope of this paper for several reasons. 
 First, we would need to use model atmospheres and temperature-color transformations suited for the proper helium range and associated peculiar composition of 2P stars to proceed from the theoretical to the observational plane. This has previously been done up to  a certain extent \citep[see, e.g.,][]{Sbordone11,Milone13,Cassisi13}, but not for the large abundance spread predicted within the FRMS scenario because such tools are not yet available.
In addition, we did not take the effects of multiple systems (e.g., unresolved binaries or those close enough to interact) or of stellar evaporation into account in our population synthesis models. 

However, we discussed in detail how the predicted distribution and number ratios of helium-rich stars drastically change with time because of the effect of the initial helium content of stars on their evolution paths and lifetimes. While the percentage of stars with an initial helium mass fraction Y$_\mathrm{ini}$ higher than 0.35 (0.4) is 33\% (21\%) within the original sample, only 24\% (12\%) remain at 9~Gyr, and 22\% (10\%) at 13~Gyr. 
These numbers remain relatively constant in the main sequence. At this phase, the effect of the very He-rich stars that are specific to the FRMS framework (compare to the AGB scenario that does not predict stars with 
Y$_\mathrm{ini}$ higher than 0.36 - 0.38) on the observed GC HRDs is therefore expected to be relatively modest.
For the horizontal branch and the asymptotic giant branch, the theoretical percentage of helium-rich stars predicted within the FRMS framework dramatically drops with increasing GC age. 
At 13~Gyr, the HB is predicted to host only 12\% of helium-rich stars (Y$_\mathrm{ini}$ above 0.35) and 0.4\% of super helium-rich stars (Y$_\mathrm{ini}$ above 0.4). 
Therefore, this evolution phase cannot be used to distinguish between the two most commonly invoked scenarios for GC enrichment for similar ages and metallicities. Additionally, the AGB of NGC~6752 is predicted to be free of very helium-rich stars within the FRMS scenario, which is supported by the lack of very Na-rich AGB stars in this relatively old and metal-poor cluster.

Finally, we compared these predictions with the He spread derived from isochrone fitting of the MS colors in UV and optical filters obtained with HST for a handful of GCs. This led us to emphasize the difficulties of the original FRMS scenario with respect to the current results  of GC CMDs photometric analysis and to propose some alternatives to avoid strong helium enrichment in GC 2P stars.
 
\begin{acknowledgements}
We acknowledge support from the Swiss National Science Foundation (FNS) for the Project 200020-159543 ``Multiple stellar populations in massive star clusters – Formation, evolution, dynamics, impact on galactic evolution" (PI C.C.). We thank the International Space Science Institute (ISSI, Bern, CH) for welcoming the activities of ISSI Team 271 ``Massive star clusters across the Hubble Time'' (2013 - 2016, team leader C.C.). We warmly thank Eugenio Carretta for kindly providing his valuable data on NGC 6752. Finally, we are indebted to the anonymous referee, whose pertinent questions and suggestions have helped us improve the presentation of our results.

\end{acknowledgements}

\bibliographystyle{aa}
\bibliography{Chantereau16}

\end{document}